\documentclass{article}

\usepackage[utf8]{inputenc}
\usepackage[english]{babel} 
\usepackage[margin=2.54cm]{geometry}
\usepackage{authblk}

\usepackage{physics,braket,fancyhdr,amssymb,lastpage,amsmath,graphicx,sidecap,xcolor,fancyref,gensymb,adjustbox,xcolor,soul,cancel,comment,mathtools,booktabs,multirow,subfig,caption,float}
\newcommand{\leri}[1]{\left(#1 \right)}
\newcommand{\mphi}{m_{\phi}}
\usepackage{hyperref} 
\usepackage{setspace} 
\usepackage[title]{appendix}

\usepackage{color}
\definecolor{lightblue}{cmyk}{1,0.3,0,0.2}
\definecolor{purplePink}{cmyk}{0,1,0,0.2}
\definecolor{cadmiumgreen}{rgb}{0.0, 0.42, 0.24}
\definecolor{brilliantlavender}{rgb}{0.96, 0.73, 1.0}
\hypersetup{colorlinks, bookmarksnumbered, citecolor=lightblue, linkcolor=lightblue, pdfstartview=FitH, urlcolor=purplePink, linktocpage}

\makeatletter
\let\oldabs\abs
\def\abs{\@ifstar{\oldabs}{\oldabs*}}
\let\oldnorm\norm
\def\norm{\@ifstar{\oldnorm}{\oldnorm*}}
\makeatother

\usepackage[section]{placeins}   
\usepackage[para]{footmisc}

\usepackage[nolist]{acronym}
\begin{acronym}[aligator]
    \acro{DM}{Dark Matter}
    \acro{DD}{Dynamic Decoupling}
    \acro{SM}{Standard Model}
    \acro{EP}{Equivalence Principle}
    \acro{CL}{Confidence Level}
    \acro{ULE}{Ultra Low Expansion}
\end{acronym}

\begin{document}
\title{Constraining Rapidly Oscillating Scalar Dark Matter Using Dynamic Decoupling}

\author[1]{Shahaf Aharony}
\author[2]{Nitzan Akerman}
\author[2]{Roee Ozeri}
\author[1]{Gilad Perez}
\author[1]{Inbar Savoray}
\author[2]{Ravid Shaniv}
\affil[1]{Department of Particle Physics and Astrophysics}
\affil[2]{Department of Physics of Complex Systems}
\affil[ ]{Weizmann Institute of Science, Rehovot, Israel 7610001}
\date{\vspace{-5ex}}

\maketitle	

\begin{abstract}
We propose and experimentally demonstrate a method for detection of light scalar \ac{DM}, through probing temporal oscillations of fundamental constants in an atomic optical transition. Utilizing the quantum information notion of \ac{DD} in a table-top setting, we are able to obtain model-independent bounds on variations of $\alpha$ and $m_e$ at frequencies up to the MHz scale. We interpret our results to constrain the parameter space of light scalar \ac{DM} field models. We consider the generic case, where the couplings of the \ac{DM} field to the photon and to the electron are independent, as well as the case of a relaxion \ac{DM} model, including the scenario of a \ac{DM} boson star centered around Earth. Given the particular nature of \ac{DD}, allowing to directly observe the oscillatory behaviour of coherent \ac{DM}, and considering future experimental improvements, we conclude that our proposed method could be complimentary to, and possibly competitive with, gravitational probes of light scalar \ac{DM}. 
\end{abstract}
\vspace{5ex}

\acresetall
{\bf Introduction.  }
The 'Missing Mass' problem is one of the most fundamental questions in modern physics~\cite{Bertone:2016nfn}.
Although particle \ac{DM} at the electroweak scale is a highly motivated solution~\cite{Jungman:1995df}, no discovery of such \ac{DM} was made to date~\cite{Bertone:2004pz,daSilva:2017swg,Aprile:2018dbl}. Another intriguing possibility is that of a sub-eV scalar \ac{DM} field, coherently oscillating to account for the observed \ac{DM} density (e.g.~\cite{Arvanitaki:2014faa,Banerjee:2018xmn,Graham:2015ifn}). A coupling between the coherent \ac{DM} candidate and the \ac{SM} particles, would result in temporal oscillations of fundamental constants, such as the fine-structure constant and the electron's mass \cite{Safronova:2017xyt, Roberts:2017hla, Derevianko:2013oaa, Derevianko:2016vpm, Arvanitaki:2014faa,Graham:2015ifn}. Here we propose and experimentally demonstrate a method probing this \ac{DM} signature in an atomic optical transition, at a bandwidth ranging from few Hz to the MHz range. This range, corresponding to light scalar \ac{DM} field which is coherently oscillating at these frequencies, has been a blind spot for current experimental measurements of time variations of fundamental constants (e.g.~\cite{Arvanitaki:2015iga,VanTilburg:2015oza,Geraci:2018fax}), despite being theoretically motivated (e.g. \cite{Dobrich:2015xca,Banerjee:2018xmn}). Our proposal uses a table-top setting and utilizes the quantum information notion of \ac{DD}~\cite{viola1999dynamical,kotler2011single} to amplify the desired signal within this uncovered bandwidth in a noisy environment\footnote{We acknowledge early discussions on this idea with Andrei Derevianko and Shimon Kolkowitz.}.\\ 

{\bf Rationale.  }
For a scalar field $\phi$ which couples to the electromagnetic field strength $F_{\mu \nu}$ and to the electron~$e$ as~\cite{Arvanitaki:2015iga,Flacke:2016szy}:
\begin{align}
    \mathcal{L}_\text{int} &\supset  \frac{g_{\phi \gamma}}{4}\phi F^{\mu \nu} F_{\mu \nu}  - g_{\phi e}\phi\bar{e} e \,, 
\end{align}
the mass of the electron $m_{e}$ and the fine-structure constant $\alpha$ will be modified with respect to their \ac{SM} values as
\begin{align}
    m_{e} &= m_e^\text{SM}+\delta m_e\,,  &
    \delta m_e &= g_{\phi e}\braket{\phi\leri{t,\vec{x}}}\,,\nonumber\\
    \alpha &= \alpha^\text{SM} + \delta \alpha\,, &
    \delta \alpha &=  g_{\phi \gamma}\braket{\phi\leri{t,\vec{x}}} \alpha^\text{SM} \,.
\end{align}
If $\phi$ is a \ac{DM} candidate, it is expected to oscillate as~\cite{Arvanitaki:2015iga}
\begin{align}
    \langle \phi\leri{t,\vec{x}}\rangle &\simeq \frac{\sqrt{2\rho_\text{DM}}}{m_{\phi}}\cos(m_{\phi}\leri{t-\vec{v}\cdot \vec{x}+\ldots})\,,
\end{align}
where $\rho_\text{DM}$ is the \ac{DM} density, $m_\phi$ is the mass of the candidate and $\vec{v}$ is its velocity relative to Earth.
Therefore, given experimental results, we can set bounds on the couplings of the \ac{DM} candidate  
\begin{align}
    g_{\phi e} & \leq d_{m_e} \mphi \frac{m_{e}}{\sqrt{2 \rho_{\text{DM}}}}\,, &
    g_{\phi \gamma} & \leq d_\alpha \mphi \frac{1}{\sqrt{2 \rho_{\text{DM}}}},~\label{generic_eqs}
\end{align}
where $d_{m_e} \equiv \leri{\frac{\delta m_e}{m_{e}}}^\text{UB}$ and $d_\alpha \equiv \leri{\frac{\delta \alpha}{\alpha}}^\text{UB}$ are upper bounds extracted from experimental measurements. For a generic \ac{DM} candidate, $g_{\phi e}$ and $g_{\phi \gamma}$ are independent. 

Experimental bounds for $d_{m_e}$ and $d_\alpha$ at a specific temporal modulation frequency $\nu$ corresponding to the scalar mass $m_{\phi}$, can be obtained by monitoring oscillations of an atomic optical transition frequency that depends on $\alpha$ and on $m_{e}$, when compared to a frequency reference that depends differently on these parameters.  However, these oscillations might be overshadowed by the noisy experimental environment. In order to amplify the desired signal while mitigating undesired noise, we propose to use \ac{DD} \cite{viola1999dynamical}.\\

{\bf Measuring Temporal Oscillations of Fundamental Constants Using \ac{DD}.  }
\ac{DD} is a notion that utilizes the application of a known time-dependent Hamiltonian $\mathcal{H}\left(t\right)$ on an open quantum system, in order to alter the effect of the environment on a specific sub-system. From a metrological point of view, $\mathcal{H}\left(t\right)$ functions as a spectral filter, screening the evolution of this sub-system outside of, and enhancing it in, an engineered spectral window.

Our experimental proposal relies on the comparison of the optical frequency of a trapped ion's optical clock transition to a narrow-linewidth laser locked to an ultra-stable cavity, and placing bounds on the amplitude of the ion-laser relative AC frequency shift at frequency $\nu$, denoted as $\Delta f\left(\nu\right)$. First, an equal superposition is created between the ion's ground and excited state, using a laser $\frac{\pi}{2} $ pulse. Next, laser $\pi$ pulses periodically rotate the ion's state around some equatorial axis on the Bloch sphere, at a chosen modulation frequency $\nu_{m}$. This modulation both filters out variations of $\Delta f \left(\nu\ne \nu_{m}\right)$ and enhances the signal from $\Delta f \left(\nu=\nu_{m}\right)$. The modulation frequency $\nu_{_m}$ is scanned, and bounds can be placed on the frequency oscillations at each $\nu_{m}$ value. A detailed description of the above method can be found in \cite{shaniv2017quantum}.

The ion's transition frequency shift is proportional to the change in the Rydberg constant $R_{\infty}\propto\alpha^{2}m_{e}$, and therefore the relative frequency change of the ion due to variations in $\alpha$ and $m_{e}$ is~\cite{Wcislo:2016}
\begin{equation}
    \frac{\delta f_{\mathrm{ion}}\left(\nu\right)}{f_{\mathrm{ion}}}=2\frac{\delta \alpha\left(\nu\right)}{\alpha}+\frac{\delta m_{e}\left(\nu\right)}{m_e}.
\end{equation}
In contrast, the laser's frequency shift is inversely proportional to the distance between the cavity mirrors $r_0$, which is proportional to the Bohr radius $a_{0}\propto\left(m_{e}\alpha\right)^{-1}$~\cite{Geraci:2018fax}. Therefore, the relative frequency change of the laser would depend on $\alpha$ and $m_{e}$ as
\begin{equation}
    \frac{\delta f_{\mathrm{laser}}\left(\nu\right)}{f_{\mathrm{laser}}}=\left(\frac{\delta \alpha\left(\nu\right)}{\alpha}+\frac{\delta m_{e}\left(\nu\right)}{m_{e}}\right)\times F\left(\nu\right).
\end{equation}
Here, $F\left(\nu\right)$ denotes a frequency-dependent response of the laser frequency to the change in $a_{0}$ at a specific signal frequency $\nu$. At frequencies much lower than the cavity's lowest mechanical mode and optical linewidth, $r_0$ follows the change in $a_{0}$, and the laser changes its frequency accordingly, meaning $F\left(\nu=\mathrm{low}\right)\rightarrow 1$. At frequencies much higher than the cavity's linewidth and the ratio between the speed of sound in the cavity $v_{\mathrm{sound}}$ and $r$, the cavity's mechanical response to the variations in $a_{0}$ is low-pass filtered, and in addition, the laser cannot follow the cavity's instantaneous frequency due to the finite lifetime of a photon in the cavity. Therefore, the laser frequency response to variations in $a_{0}$ is further reduced, and $F\left(\nu=\mathrm{high}\right)\rightarrow0$. Assuming $f_{\mathrm{cavity}}\approx f_{\mathrm{ion}}=f_{0}$, we obtain
\begin{equation}
    \frac{\Delta f \left(\nu\right)}{f_{0}}=\frac{\delta f_{\mathrm{ion}}\left(\nu\right)-\delta f_{\mathrm{laser}}\left(\nu\right)}{f_{0}}\nonumber \\
    =\left(2-F\left( \nu\right)\right)\frac{\delta \alpha}{\alpha}+\left(1-F\left( \nu\right)\right)\frac{\delta m_e}{m_e}, \label{sensitiviy alpha vs me}
\end{equation}
leading to the conclusion that at low frequencies only $\alpha$ variation is detectable, whereas at high frequencies variations in both constants may be observed.

Below we present bounds on \ac{DM} obtained from of a proof-of-principle experiment, in which we used a laser at $674$~nm locked to $r=0.1$~m long, high-finesse (300,000) \ac{ULE} optical Fabri-P\'erot cavity, with a $4.5$~kHz linewidth. This laser frequency matched the clock dipole-transition $5S_{\frac{1}{2}}\leftrightarrow 4D_{\frac{5}{2}}$ of a single $^{88}\text{Sr}^{+}$ ion, on which the \ac{DD} sequence was applied. Here, $\nu_{m}=1013$~Hz was chosen. The difference between the superposition phase and the laser phase was mapped onto the populations of the ground and excited states by applying a final interrogation laser pulse and scanning its phase between $0$ and $2\pi$. Assuming no synchronization between the control optical pulses modulation phase and the optical frequency oscillation at $\nu_{m}$, a bound for the transition frequency modulation amplitude was readily inferred from the deviation of the resulting Ramsey fringe amplitude from $0.5$ (see Appendix~\ref{filter_function_appendix}). More details about the experimental parameters and setup can be found in \cite{shaniv2017quantum}.

\begin{figure}[h]
\centering
\includegraphics[scale=0.3]{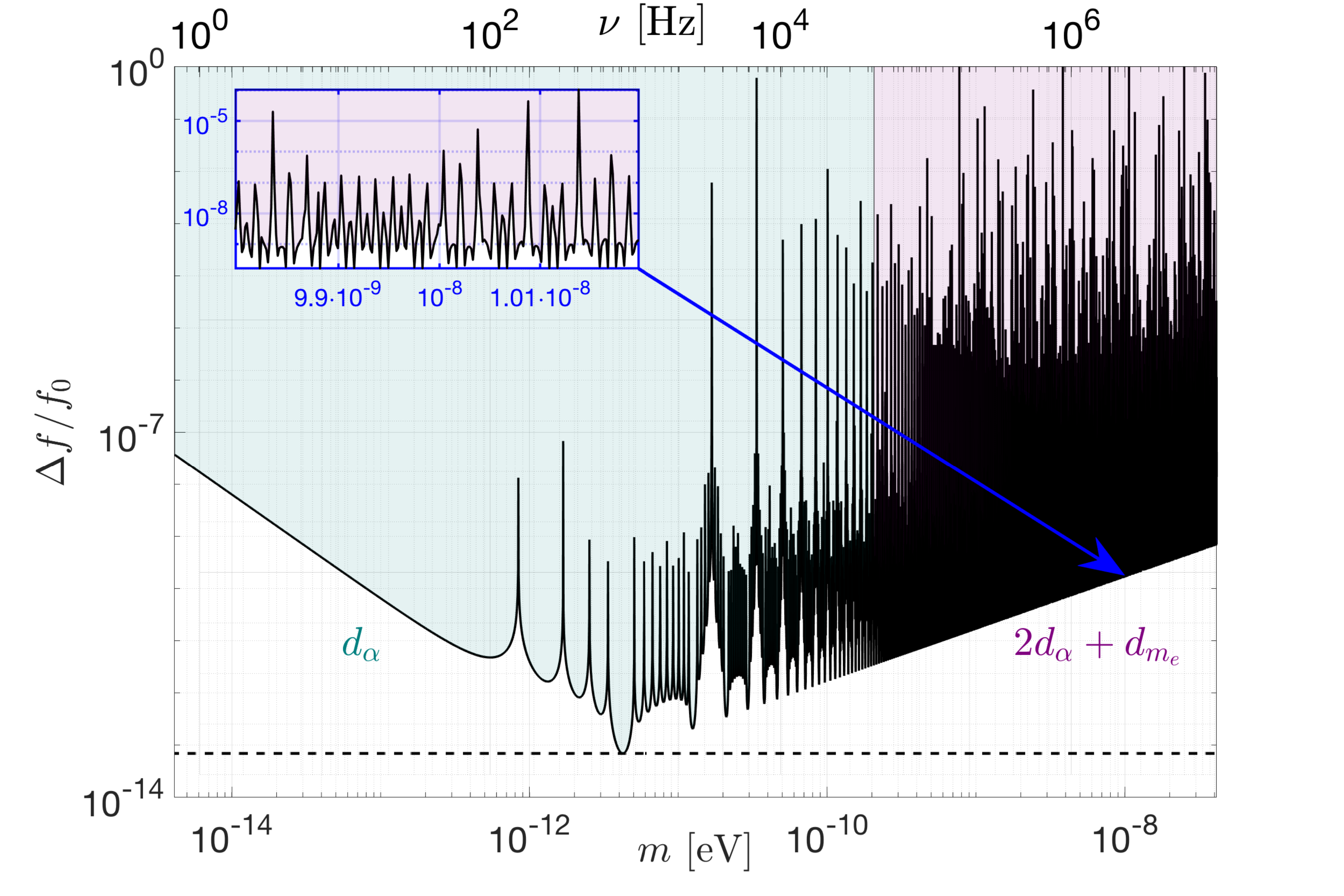}
\caption{Current bound on the relative modulation of the transition frequency from a \ac{DD} experiment, placed at 95\%~\ac{CL}. The dashed line marks the current sensitivity reach, corresponding to scanning over $\nu_m$. The inset is a magnified view of $m\sim 10^{-8}$eV.}
\label{current_measurement}
\end{figure}

The upper bound for the relative frequency modulation amplitude at different values of $\nu$ is given in Fig.~\ref{current_measurement}. The sharp peaks appear at frequencies where the experimental \ac{DD} sequence loses sensitivity (see Appendix~\ref{filter_function_appendix}). The best sensitivity is obtained for $\nu=\nu_{m}$, and by scanning this frequency high sensitivity can be maintained for the entire scan range. The dashed line in Fig.~\ref{current_measurement} is the expected bound obtained from performing the experiment proposed above for different $\nu_{m}$ and measuring the same Ramsey fringe amplitude as for $\nu_{m}=1013$ kHz. The bound on $\delta f/f_0$ corresponds to bounds on $d_\alpha$ or $2d_\alpha+d_{m_e}$, as explained above and shown in the plot. While stricter constraints on these parameters already exist in the literature~\cite{Hees:2018fpg,Wcisloeaau4869,VanTilburg:2015oza,PhysRevLett.113.210801}, our bound is currently the only one directly constraining temporal oscillations of $\alpha$ and $m_e$ in the 10~Hz-MHz frequency range. Beyond this, model-independent, experimental proof of principle, we are interested in the implications of our measurements to models of light scalar \ac{DM}.

The bound in Fig. \ref{current_measurement} is mostly limited by experimental imperfections, and can be improved by at least two orders of magnitude. We also note that in the current setup, the signal was encoded in the coherence of the ion's superposition, or alternatively the amplitude of the Ramsey fringe \cite{shaniv2017quantum}. Since both the signal and the experimental imperfections (e.g. $\pi$ pulse fidelity) tend to decrease the fringe amplitude, the bound would be ultimately limited by the experimental apparatus. However, in the case of large enough quality factor of the $\delta m_{e}$ and $\delta\alpha$ oscillations, it would be useful to synchronize different experiment realizations via an external clock, such that for a specific $\nu_{m}$, different experimental realizations would measure signal oscillations with a known phase difference \cite{schmitt2017submillihertz,boss2017quantum}. This would allow to infer the signal amplitude from the final superposition phase, separating it from the coherence of our atom. \\

\begin{figure}[]
\centering
\subfloat[][]{\includegraphics[scale=0.4]{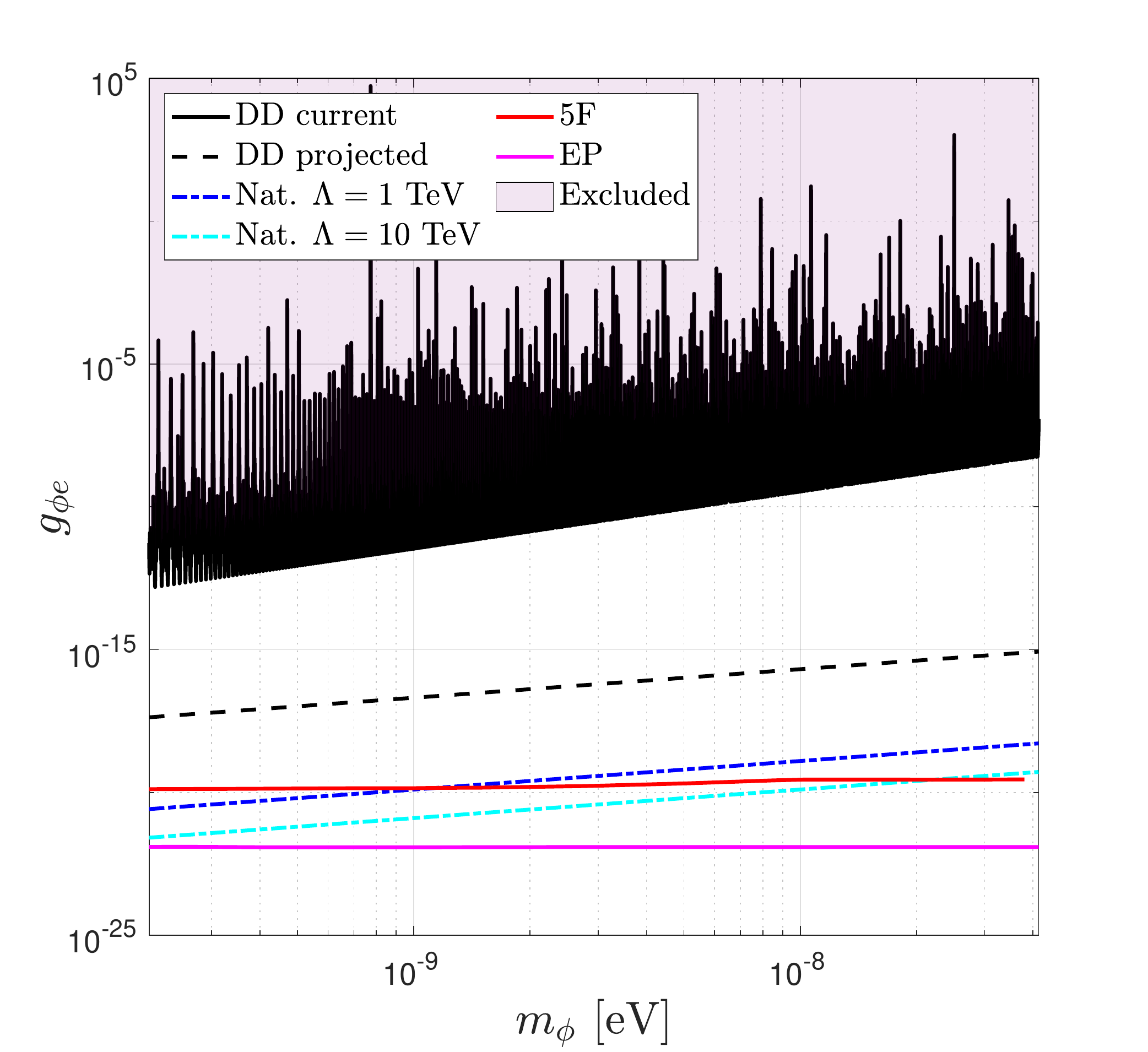}\label{gphie all}}
\subfloat[][]{\includegraphics[scale=0.4]{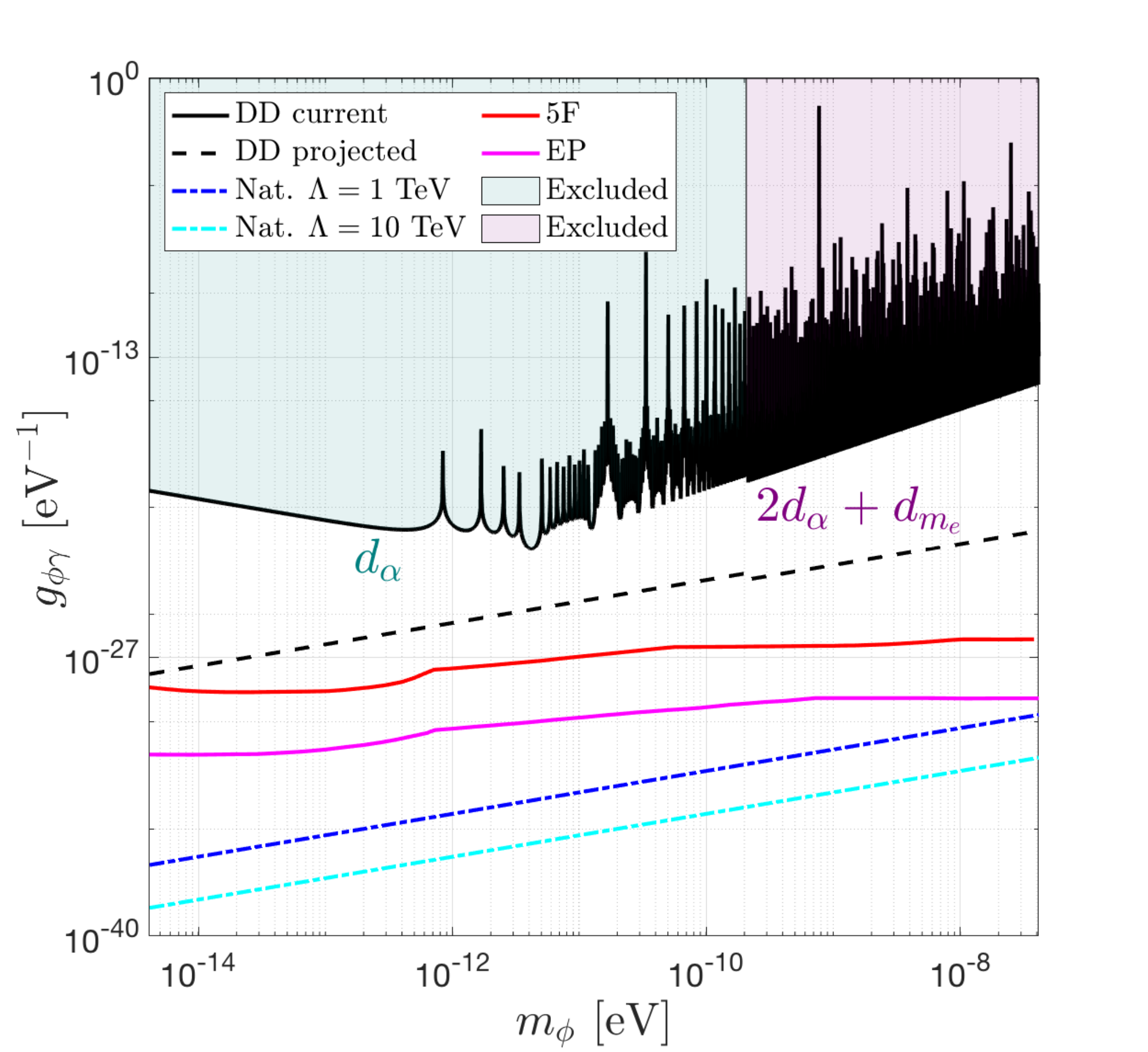}\label{gphigamma all}}
\qquad
\subfloat[][]{\includegraphics[scale=0.4]{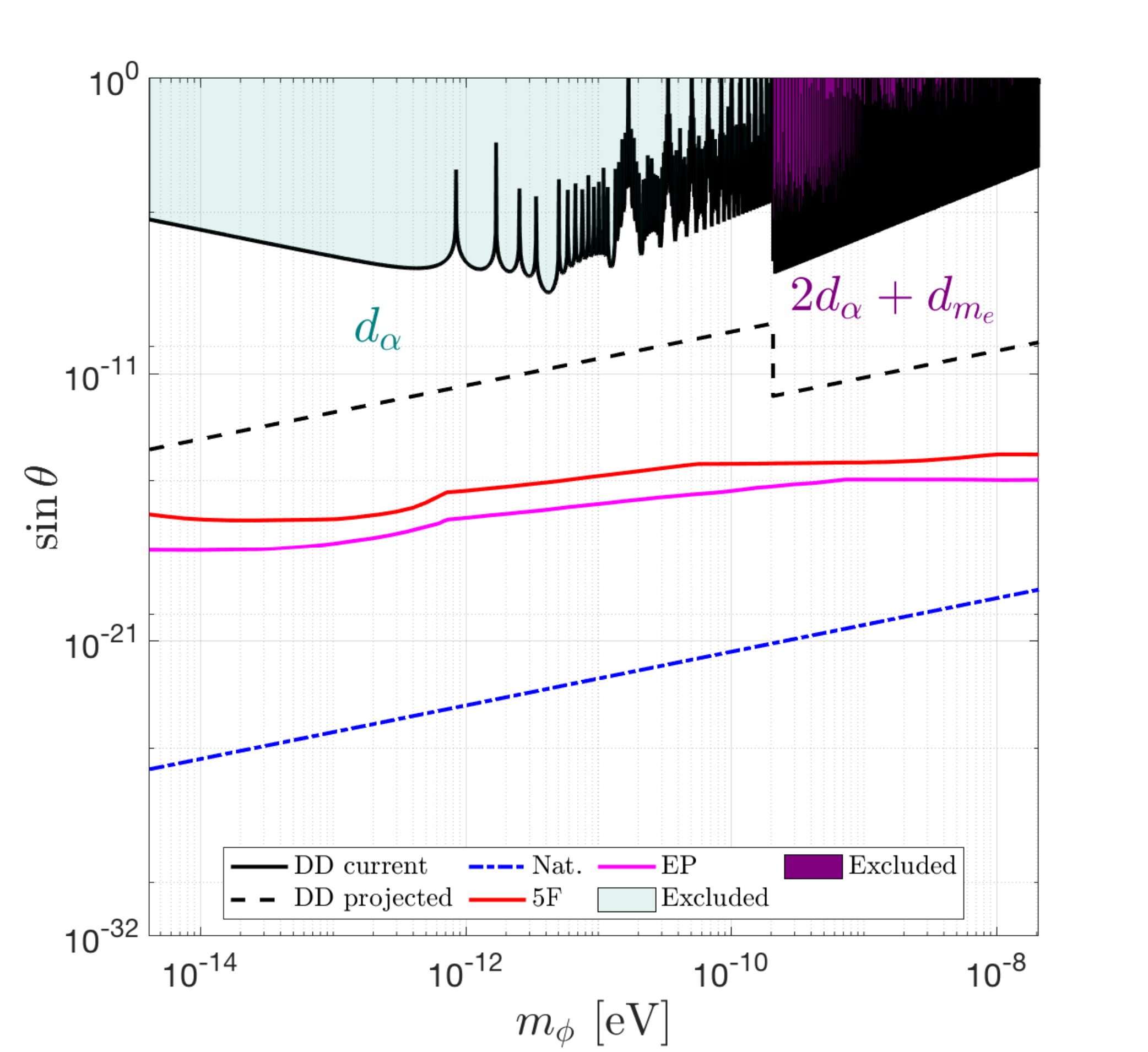}\label{sin theta nat}}
\caption{Bounds on the parameter space of light scalar \ac{DM} corresponding to the observed \ac{DM} density near the sun. The bounds on the couplings of a generic \ac{DM} candidates are shown in (a) and (b). The bounds on the mixing angle of a relaxion \ac{DM} are presented in (c). \textit{Black} -- current and projected bounds from DD experiments at 95\% CL. \textit{Red} -- Bounds from fifth force experiments~\cite{Adelberger:2003zx}. \textit{Magenta} --  \ac{EP}-tests bounds taken from~\cite{Hees:2018fpg}. \textit{Dash-dotted} -- Bounds from Naturalness.}
\label{results_models}
\end{figure}

{\bf Bounds on Light Scalar Dark Matter from \ac{DD} Experiments.  }
Using these results, we obtain upper limits on the values of $g_{\phi e}$ and $g_{\phi \gamma}$ at 95\%~\ac{CL}, and present them in Fig.~\ref{gphie all} and Fig.~\ref{gphigamma all}, respectively. The background \ac{DM} density is assumed to be $\rho_\text{DM}=\rho_{\text{DM}_{\odot}}= 3.1 \cdot 10^{-6}\text{~eV}^4$, which is the local \ac{DM} density around the sun~\cite{Salucci:2010qr}. As shown in Eq.~\ref{sensitiviy alpha vs me}, the measurement is sensitive to variations in $\alpha$ in the entire range, but is not sensitive to variations in $m_e$ at low frequencies. For our analysis, we assumed a sharp transition between $d_\alpha$ sensitivity and $d_{m_e}+2d_\alpha$ sensitivity at $\nu=50$~kHz, namely $F(\nu) = \Theta (\nu-50 \text{ kHz})$, where $\Theta$ is the Heaviside step function. The step frequency $\nu_{\text{step}}=50$ kHz is the ratio between the speed of sound in our cavity spacer $v\approx 5 \frac{\text{km}}{\text{s}}$ and the cavity length $r\approx 0.1$ m. The optical linewidth is an order of magnitude lower. The black dashed line is the projected sensitivity for the proposed method, corresponding to scanning the experimental modulation frequency $\nu_{m}$. We compare our results with the current bounds obtained from experiments testing deviations from gravity (\ac{EP} and fifth force) and Naturalness - both are explained below.

We would like to further interpret our results in accordance to the relaxion~\cite{Graham:2015cka} model, which was recently shown to be a viable \ac{DM} candidate~\cite{Banerjee:2018xmn}. The interactions of a relaxion \ac{DM} with the \ac{SM} fields are mediated through its mixing with the Higgs, and thus the corresponding couplings are no longer independent.  The couplings of the relaxion to the electron and to the photon are given by~\cite{Flacke:2016szy}
\begin{align}
g_{\phi e}&=Y_e\sin{\theta}\,,\nonumber\\
g_{\phi \gamma}&=-\frac{ \alpha_0 \sin \theta }{2\pi v}\left| A_W(\tau_W)+\sum_{\rm fermions} N_{c,f} Q_f^2 A_F(\tau_f)\right|\label{gphigamma with mixing} \,,
\end{align}
where $Y_e$ is the Yukawa coupling of the electron to the Higgs ($h$), $\theta$ is the mixing angle between the relaxion and the Higgs and $\tau_x = m^2_h / 4 m_x^2$. $A_F(\tau)$ and $A_W(\tau)$ are defined in~\cite{Flacke:2016szy} and calculated accordingly. The upper bound on $\Delta f/f_0$ can then be used to exclude the region in the $\mphi - \sin(\theta)$ parameter space corresponding to
  \begin{align}
     \sin \theta \leq  \left. \leri{\frac{\Delta f}{f_0}}^\text{UB}\frac{m_{\phi}}{\sqrt{2\rho_{\text{DM}_{\odot}}} \leri{\frac{\sqrt{2}\kappa_e}{v}\left(1-F\right)-1.12 \cdot 10^{-14}\left(2-F\right)}}  \right. \label{higgs_portal_eq}\,,
\end{align}
where $\kappa_e \equiv Y_e/Y_e^\text{SM}$. Analyzing the results in Fig.~\ref{current_measurement} for relaxion \ac{DM}, we obtain the appropriate upper limit on $\sin{\theta}$, assuming $\kappa_e=1$, and present it in Fig.~\ref{sin theta nat}. Note that currently the upper bound on $\kappa_e$ is at $6.1\cdot 10^{2}$~\cite{Dery:2017axi}, which would yield a stronger constraint.
The analysis presented here can also be modified to apply to other Higgs Portal \ac{DM} scenarios~\cite{Piazza:2010ye}.

\begin{figure}[] 
\centering
\subfloat[][]{\includegraphics[scale=0.55]{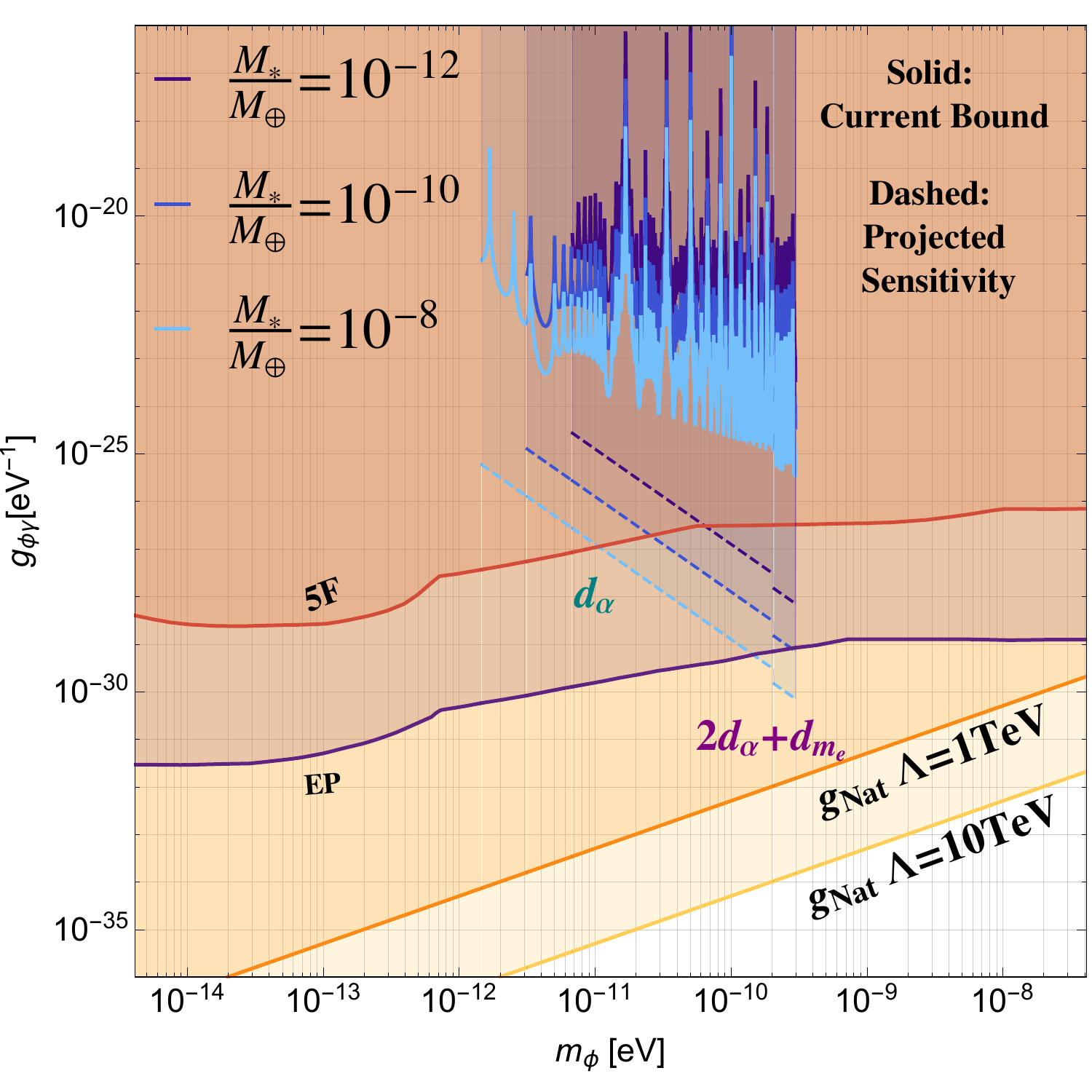}\label{g_gamma_relaxion_star}}
\subfloat[][]{\includegraphics[scale=0.55]{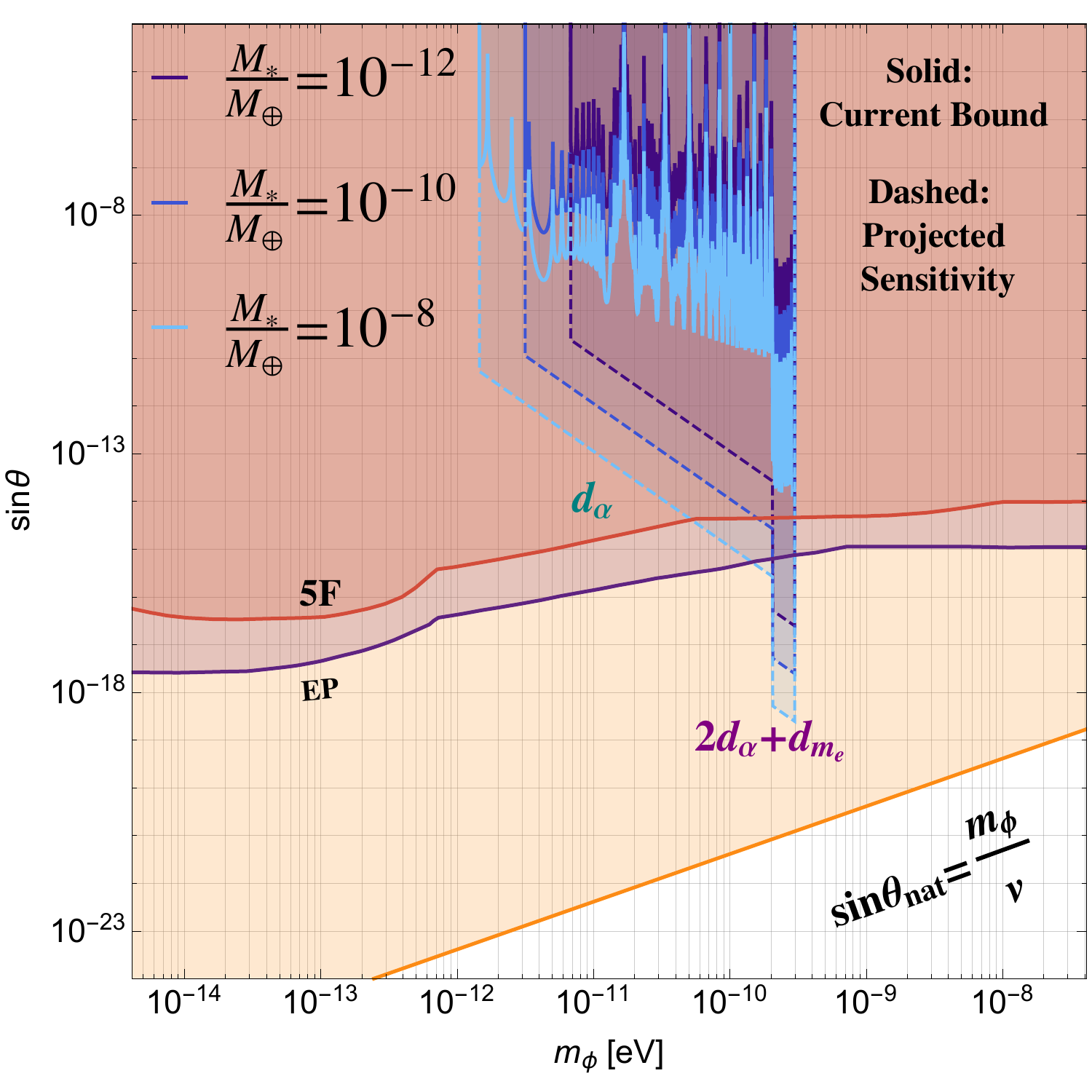}\label{sin_relaxion_star}}
\caption{Bounds for a relaxion star centered around Earth.}
\label{relaxion star bounds}
\end{figure}

Relaxions, being light scalar fields, can form what are known as boson stars~\cite{Colpi:1986ye}. Such stars could either pass through Earth, or be bound to its gravitational potential. The latter scenario will increase the \ac{DM} density around Earth at all times and, accordingly, the signal measured in our proposed experiment~\cite{Banerjee:2019}. The mass of the relaxion star is constrained by local measurements of gravitational acceleration~\cite{Adler:2008rq}, and should satisfy $M_\star \lesssim 10^{-8} M_\oplus$, where $M_\oplus$ is the mass of the Earth. Setting the radius of the star to support the balance between kinetic and gravitational energy (see Appendix~\ref{relaxion_stars_appendix}), the density of the star $\rho_\star$ follows
\begin{align}
    \rho_\star &= \frac{81\,m^6}{32\pi\,M_P^6}\,M_\oplus^3\,M_\star\,,
\end{align}
where $M_P$ is the Planck mass. Using the above density profile as the \ac{DM} density, combined with the bounds shown in Fig.~\ref{current_measurement}, we obtain upper limits on $g_{\phi\gamma}$ and $\sin\theta$. The bounds for the scenario of a relaxion star around Earth are shown in Figs.~\ref{g_gamma_relaxion_star} and~\ref{sin_relaxion_star}, respectively. We consider the mass region corresponding to both $R_\star\geq 10\cdot R_\oplus$, justifying the point-like external mass approximation (see Appendix~\ref{relaxion_stars_appendix}), and $\rho_\star \geq \rho_{\rm{DM}_{\odot}}$, allowing for a stronger sensitivity compared to the background density scenario.

There are other constraints on the parameter space of a light scalar \ac{DM}. The first arises from \ac{EP} or fifth force experiments \cite{Hees:2018fpg}, designated to detect deviations from gravity. The bounds related to \ac{EP}-tests presented here are based on those given in~\cite{Hees:2018fpg}. Fifth force experiments are specifically sensitive to inverse square law-violating Yukawa interactions. The bound presented here is based on the $95\%$-\ac{CL} constraints from \cite{Adelberger:2003zx} and the analysis is equivalent to that carried out in~\cite{Arvanitaki:2015iga}. Although the bounds we obtain from \ac{DD} are currently weaker than those set by the experimental tests of gravity, they are different in essence. The constraints resulting from \ac{DD} experiments are directly related to the temporal variations of $\alpha$ and $m_e$, whereas fifth force and \ac{EP}-tests are only sensitive to time independent, or very slow, shifts. This difference would be important in the case of a discovery of a rapidly oscillating scalar \ac{DM}. While gravitational tests could only indicate a possible candidate, our proposed method could also directly observe its oscillatory nature, and thus positively identify it as a coherent \ac{DM} field. In addition, it can be seen that for some region of the parameter space, our future-projected bounds could become competitive with those of gravitational tests for the scenario of a \ac{DM} star, even before fully exhausting the experimental improvements suggested above. It is also worth noting that the relation between the observables of the different experiments and the \ac{DM} parameters is model dependent, and can be modified in the case of a non-linear interaction to yield a different interplay between gravity-related and atomic bounds~\cite{Hees:2018fpg}.  

Another set of constraints comes from Naturalness. Since $\phi$ is a scalar field, its mass parameter is sensitive to radiative corrections resulting from its interactions. To maintain Naturalness, we require these quantum corrections to be small compared to the bare mass $\delta m^2_{\phi_\text{1-loop}}{\ll}m^2_{\phi_\text{bare}}$. For a scalar field with the interaction terms describes above, this would imply~\cite{DIMOPOULOS1996105,Graham:2015ifn,Arvanitaki:2015iga}
\begin{align}
\abs{g_{\phi e}}&\ll \frac{4\pi m_{\phi}}{\Lambda}\,, &
\abs{g_{\phi \gamma}}&\ll \frac{16\pi m_{\phi}}{\Lambda^2}\,.
\end{align}
For the case of a relaxion \ac{DM}, the constraint reduces to $\sin\theta\leq \frac{m_\phi}{v}$~\cite{Frugiuele:2018coc}.\\

{\bf Conclusion.  }
Rapidly oscillating scalar \ac{DM} field is a well-motivated scenario, but currently lies in a blind spot of existing experimental searches sensitive to coherent oscillations of fundamental constants. In this letter, we have proposed a new experimental probe of light scalar \ac{DM}, utilizing the method of \ac{DD} in a table-top setting. Using a proof-of-concept experimental measurement, we have obtained model-independent bounds on the temporal oscillations of both $m_e$ and $\alpha$ at frequencies up to MHz scale. Consequently, we were able to set upper limits on the couplings of a generic coherent \ac{DM} candidate. We have also interpreted the results for the case of relaxion \ac{DM}, including the scenario of a relaxion star centered around Earth, for which our constraints are significantly tightened. As an experimental outlook, we believe that the bounds presented here can be improved significantly in two ways. First, the modulation frequency $\nu_{m}$ can be scanned and therefore a lower bound can be achieved for a range of frequencies. Second, instead of the superposition coherence, its phase shift can be measured by synchronizing different experimental realizations, separating the desired signal from unwanted experimental imperfections. Therefore, our proposed method could be an important tool for studying light scalar \ac{DM}, not only directly probing its oscillatory nature, but also possibly setting constraints that would be competitive with fifth-force and \ac{EP} tests in the future. Additional measurements covering complementary parts of the spectrum have been recently concluded and their reports are in preparation \cite{JunYe2019, Budker2019}.\\ 

{\bf Acknowledgements.  } We would like to thank A. Derevianko, S. Kolkowitz and D. Budker for useful discussions. We thank D. Budker  and Y. Nir for comments on the manuscript. We are grateful to J. A. Eby and H. Kim for the assistance and the advice regarding relaxion stars. R. O and R. S. acknowledge support by the Crown Photonics Center, ICORE-Israeli excellence center Circle of Light, The Israeli Science Foundation, the Israeli Ministry of Science Technology and Space and the European Research Council (consolidator grant 616919-Ionology). The work of GP is supported by grants from the BSF, ERC, ISF; the work of RO and GP is jointly support by the Minerva Foundation, and the Segre Research Award. 

\appendix
\section{Filter function}\label{filter_function_appendix}
We assume an oscillating atomic angular frequency in the form of
\begin{equation}
    \delta\left(t,\xi\right)=2\pi f_{0}\sin{\left(2\pi f_{s} t+\xi \right)}.
\end{equation}
Applying optical $\pi$ pulses in a repeating unit cell of
\begin{center}
[wait time $\tau$]--[$\pi$ pulse]--[wait time $\tau$]    
\end{center}
results in a phase modulation kernel of the form 
\begin{equation}
    f\left(t,\tau,n\right)=\mbox{rect}\left(\frac{t}{2n\tau}\right)\left[\Theta\left(t\right)+2\sum_{k=1}^{\infty}\left(-1\right)^{k}\Theta\left(t-\left(2k-1\right)\tau\right)\right],
\end{equation}
where $\Theta$ is the Heaviside step function, $n$ is the number of pulses and $\mbox{rect}$ is a rectangular window function nulling the modulation at $t<0$ and $t>2n\tau$. The resulting superposition phase is therefore
\begin{equation}
       \phi\left(t,\tau,n,\xi\right)=\int_{-\infty}^{\infty}f\left(t,\tau,n\right)\delta\left(t,\xi \right) dt=4\frac{f_{0}}{f_{s}}\times\frac{\cos\left(2\pi f_{s}n\tau+\xi+n\frac{\pi}{2}\right)\sin\left(2\pi f_{s}n\tau-n\frac{\pi}{2}\right)\sin^{2}\left(\frac{2\pi f_{s}\tau}{2}\right)}{\cos\left(2\pi f_{s}\tau\right)}.
\end{equation}
This would be the signal corresponding to a phase estimation experiment. In the experimental bound presented in this work, the theoretical fringe amplitude $A$ takes the form of
\begin{equation}
    A\left(t,\tau,n\right)=0.5\left|\left<\cos\left(\phi\left(t,\tau,n,\xi\right)\right)\right>_{\xi}\right|,
\end{equation}
where $\left<\cdot\right>_{\xi}$ denotes averaging over $\xi$'s sampled from a uniform distribution between $0$ and $2\pi$. The resulting contrast is given by
\begin{equation}
    A\left(t,\tau,n\right)=0.5 \left|J_{0}\left(4\frac{f_{0}}{f_{s}}\times\frac{\sin\left(2\pi f_{s}n\tau\right)\sin^{2}\left(\frac{2\pi f_{s}\tau}{2}\right)}{\cos\left(2\pi f_{s}\tau\right)}\right)\right|,
\end{equation}
where $J_{0}$ is the zeroth Bessel function of the first kind. This function was used in the analysis of Fig.~\ref{current_measurement} for bounds on $f_{0}$.

\section{Relaxion Star}\label{relaxion_stars_appendix}
A boson star is typically supported by a balance of forces, between the (repulsive) kinetic energy of the constituent scalars, and the (attractive) gravitational interaction. If the mass $M_\star$ of the star is much smaller than $M_{ext}$, the mass of the external gravitational source the star is bounded to, then these forces will be balanced only when the radius of the star is~\cite{Banerjee:2019}
\begin{equation}
 R_\star \approx \frac{2M_P^2}{3\,m^2\,M_\oplus}\,,
\end{equation}
where we have assumed the gravitational potential $V_\text{ext}$ of the external mass $M_\text{ext}$ can be approximated as a point, $V_\text{ext} = -G\,M_\text{ext}/r$. This will be appropriate if $R_\star \gtrsim R_{ext}$. In the regime of interest, the external mass will be the Earth, $M_\text{ext} = M_\oplus$. The density of the relaxion star will then be given by
\begin{equation}
 \rho_\star = \frac{3\,M_\star}{4\pi\,R_\star^3}\,.
\end{equation}

\bibliographystyle{JHEP}
\providecommand{\href}[2]{#2}\begingroup\raggedright\endgroup

\end{document}